# Information Latency: Theory and Economic Consequences

**Gaurav Subedi[1] and Ratna K. Shrestha[2,*]**

[1] RALP Technologies, Vancouver, Canada

[2] Sauder School of Business, University of British Columbia, Vancouver, Canada

**Abstract**

Many economic decisions rely on observations that predate the decision time. This paper introduces *information latency*—the elapsed time between a decision and the most recent observation of an evolving payoff-relevant state—as a distinct source of imperfect information. Unlike classical sources of information friction, latency generates irreducible uncertainty even with exact observations and rational expectations. For sufficiently regular continuous-time Markov processes, conditional variance increases locally with latency; under the Ornstein–Uhlenbeck benchmark, it increases at a decreasing rate toward a finite bound. Following a rare-state observation, however, uncertainty can peak at an intermediate latency before declining. Neither the average latency nor the average update frequency fully characterizes information quality: the regularity of information arrival matters independently of the mean interval. Under a CARA-normal benchmark, latency-induced uncertainty generates a risk premium. With heterogeneous latency, the lowest-risk-adjusted-cost intermediary wins the business, while competition drives the price toward the second-lowest risk-adjusted cost. In a stylized contract-renewal setting, positive latency can be second-best optimal by limiting costly reclassification when direct contractual commitment is unavailable. The framework applies to credit, insurance, financial markets, and operational decisions.

* Corresponding author: ratna.shrestha@ubc.ca or shrestha.rk11@gmail.com

## 1. Introduction

Many economic decisions rely on information collected before the decision is made. Lenders evaluate borrowers using historical financial records; insurers price coverage using prior claims or medical information; and investors value firms using periodic corporate disclosures. Even when these observations are accurate, they may be imperfectly informative about current conditions because the underlying economic state continues to evolve.

This paper introduces *information latency*—the elapsed time between the decision date and the most recent observation of an evolving payoff-relevant state available to the decision maker—as a distinct source of imperfect information. Unlike asymmetric information, costly acquisition, noisy or dispersed signals, and rational inattention, latency generates uncertainty even when all available information is observed exactly and processed rationally. The baseline model treats this latency as exogenous; such latency may arise from predetermined reporting schedules, verification or processing delays, communication constraints, or institutional requirements.

This distinction separates information latency from the familiar frictions studied in information economics. Classical models emphasize asymmetric information (Akerlof, 1970; Spence, 1973; Rothschild and Stiglitz, 1976; Holmström, 1979). Other strands examine costly information acquisition (Grossman and Stiglitz, 1980), noisy or dispersed information (Veldkamp, 2006; Hellwig and Veldkamp, 2009), costly information processing and limited attention (Sims, 2003; Matějka and McKay, 2015), and dynamic learning and information aggregation through market interaction (Vives, 1993). None of these mechanisms absolutely requires the state itself to keep evolving after it is observed for the friction to exist; latency does, and isolates the uncertainty this interaction alone creates.

We formalize the mechanism in a continuous-time framework in which decision makers observe an evolving state with a fixed delay and know its true stochastic law. We show that latency generates irreducible conditional uncertainty even under rational expectations. For short delays, conditional variance increases approximately linearly with latency, at a rate determined by the state's instantaneous variance. This local expansion extends beyond diffusions to sufficiently regular continuous-time Markov processes. In the Ornstein–Uhlenbeck benchmark, conditional variance increases at a decreasing rate and converges to a finite upper bound.

The global relationship is more subtle: conditional uncertainty is higher on average when latency is longer, but this need not hold for every individual observation. Consider a lender who observes a borrower's account as delinquent—a relatively rare event. Immediately afterward, the borrower's underlying financial condition remains predictably troubled. A little later, both recovery and continued delinquency become plausible, so uncertainty peaks. Later still, as the old observation loses relevance, beliefs revert toward the typical state, and uncertainty falls again.

Thus after a rare, informative observation, uncertainty can rise, peak, and then fall as latency increases, even though it is higher on average across all possible observations.

Independent of latency itself, how regularly information arrives also matters. Two systems can update with the same average frequency yet deliver very different information quality. A satellite that revisits a location every eight days on a fixed schedule can provide more valuable information than one with the same average revisit rate but irregular gaps. This is because a randomly timed check is more likely to fall within a long gap between updates than within a short one. Thus, neither average latency nor average update frequency alone is enough to describe how good an information system really is.

We next translate these statistical results into economic consequences. Under constant absolute risk aversion, conditional normality, and linear state exposure, we derive a closed-form *latency premium* that compensates risk-averse decision makers for unresolved current-state risk—for example, the higher interest rate a lender charges, or the extra premium an insurer charges. When multiple lenders or insurers with different latencies compete for the same client, the one with the lowest risk-adjusted latency cost wins the business, but prices just below the next-lowest-cost rival's cost.

Finally, fresher information need not always improve welfare when its use undermines risk sharing. Suppose an insurer could observe a policyholder's health in real time and immediately reprice or cancel coverage the moment it worsens: the policyholder could then lose coverage exactly when it is needed most. Delaying the insurer's information—even though it makes each renewal decision less accurate—protects the policyholder by limiting the insurer's ability to act on adverse news. Building on Hirshleifer (1971) and Hendel and Lizzeri (2003), we show this formally in a stylized contract-renewal model. This does not imply that stale information is intrinsically valuable. When such a restriction can be enforced—as under the U.S. Card Act (2009) that restricts certain forms of repricing of existing balances—current information combined with that restriction weakly dominates deliberate delay.

Section 2 discusses the related literature. Section 3 presents the information structure and stochastic framework. Section 4 characterizes the statistical consequences of information latency. Section 5 derives its decision-theoretic and pricing implications. Section 6 discusses applications. Section 7 concludes with policy implications.

## 2. Related Literature

This paper relates to four main strands of literature: information economics; information processing and dynamic learning; timely disclosure and market transparency; and information freshness and Age of Information.

### *2.1 Information Economics*

The pioneering work of Akerlof (1970) shows how hidden information may cause adverse selection. Spence (1973) demonstrates how informed agents may credibly signal private information, while Rothschild and Stiglitz (1976) characterize competitive insurance markets under asymmetric information. Subsequent work (e.g., Stiglitz and Weiss, 1981) extends these ideas to credit, labor, and contract markets. In this literature, imperfect information generally arises because agents possess different private information, receive signals of different precision, or face different acquisition costs. Our framework, by contrast, requires none of these. Instead, imperfect information arises purely from the interaction between delayed observation and an evolving payoff-relevant state.

### *2.2 Information Processing and Dynamic Information*

A second strand studies how rational agents acquire, process, and use information under uncertainty. Grossman and Stiglitz (1980) show that costly information acquisition prevents prices from becoming fully revealing, while Sims (2003) introduces rational inattention. Subsequent work analyzes dispersed information, dynamic learning, and endogenous information acquisition (e.g., Veldkamp, 2006; Hellwig and Veldkamp, 2009; Zhong, 2022; and Hébert and La'O, 2023).

Other contributions examine information timing and delay in related economic and actuarial settings. Daley and Green (2012) show how anticipated future information about asset quality can induce strategic delay in trade. A related actuarial literature develops stochastic reserving methods to estimate liabilities for claims that have occurred but have not yet been reported, using historical claims and reporting patterns (Mack, 1993; England and Verrall, 2002). Whereas this literature addresses delayed reporting of past events, our framework isolates the uncertainty generated by delayed observation of an evolving state, assuming costless acquisition, rational processing, and exact observation of the state at the time it was recorded.

Filtering theory provides the closest statistical connection to our analysis. Kalman (1960) and Sinopoli et al. (2004) study estimation and prediction when a state evolves dynamically and observations are noisy or intermittent. Because our Assumption 4 rules out measurement error entirely, there is no filter gain or signal-extraction problem to solve; the relevant object is the uncertainty that accumulates between observations. A related literature in empirical macroeconomics similarly addresses reporting lags, estimating current economic conditions from partially available and staggered data releases (Giannone, Reichlin, and Small, 2008). Our focus is different: we study the economic consequences—in particular the pricing—of uncertainty generated by delayed observation itself.

### *2.3 Timely Information, Disclosure, and Market Transparency*

A well-established empirical literature links corporate disclosure and information technology to reduced information asymmetry, greater market liquidity, price efficiency and lower cost of capital (Botosan, 1997; Leuz and Verrecchia, 2000). More recent research investigates the dissemination of public information through information intermediaries (Bushee et al., 2010) and advances in electronic disclosure systems, such as SEC's EDGAR system (Goldstein, Yang, and Zuo, 2023). Collectively, these studies broadly support the view that timely information improves economic outcomes.

Timelier information need not improve welfare, however, when its use undermines risk sharing. Hirshleifer (1971) shows that early information can eliminate risk-sharing opportunities, while Hendel and Lizzeri (2003) show how contractual commitment can protect policyholders against reclassification risk. A related concern arises in financial reporting: Plantin, Sapra, and Shin (2008) show that mark-to-market accounting can generate destabilizing feedback in illiquid markets. Section 5.6 builds on this insight and derives conditions under which an exogenous observation lag can serve as a second-best substitute for restrictions on the use of newly observed information.

### *2.4 Information Freshness and Age of Information*

The communications engineering literature studies Age of Information (AoI), the time elapsed since the most recently received update was generated. Beginning with Kaul, Yates, and Gruteser (2012), this literature develops queueing models, scheduling algorithms, and transmission policies to maintain information freshness in wireless networks, sensor systems, and other applications (see Yates et al., 2021). It also examines information age under renewal updating and allows staleness costs to depend nonlinearly on age (Sun et al., 2017; Yates et al., 2021). Ornee and Sun (2021) study remote estimation of an Ornstein–Uhlenbeck process, using the nonlinear relationship between estimation error and information age to characterize sampling policies. A related refinement, the Age of Incorrect Information (Maatouk et al., 2020), measures how long a receiver's belief remains wrong, rather than simply how old it is.

Our analysis shares this statistical link between information age and uncertainty and uses standard renewal methods to characterize information age at a randomly selected decision time. Taking information timing as exogenous, we translate age-dependent variance into economic risk compensation and distinguish variability in realized age from variability in update intervals. Neither average latency nor average update frequency alone fully characterizes an information system's economic quality—a distinction we formalize in Section 4.4.2.

## 3. Model

*3.1 Information Structure*

Consider a decision maker who chooses an action at time *t*. The payoff from this decision depends on the contemporaneous value $X_t$ of an underlying payoff-relevant economic state. Depending on the application, $X_t$ may represent a borrower's creditworthiness, a firm's fundamental value, or a policyholder's risk profile, as discussed in Section 6.

Uncertainty is represented by a complete filtered probability space satisfying the usual conditions (Øksendal, 2003): $(\Omega, \mathcal{F}, \mathbb{P}, \{\mathcal{F}_t\}_{t\geq 0})$, where $\Omega$ is the sample space, $\mathcal{F}$ is the associated $\sigma$-algebra of events, $\mathbb{P}$ is the underlying probability measure, and $\{\mathcal{F}_t\}_{t\geq 0}$ denotes the filtration generated by the state process $\{X_t\}_{t\geq 0}$ through time *t*: $\mathcal{F}_t = \sigma(X_s: s \leq t)$. Under continuous observation, a decision maker acting at time *t* would condition decisions on the real-time information set $\mathcal{F}_t$.

With information latency $\tau \geq 0$, the most recent observation available at time *t* corresponds to the state $t - \tau$. The decision maker therefore conditions on the delayed information set $\mathcal{F}_{t-\tau}$. When $\tau = 0$, information is contemporaneous; and when $\tau > 0$, the available information describes an earlier realization of the evolving state. Except in Section 5.6 where $\tau$ is chosen optimally by a contracting party, $\tau$ is treated throughout as exogenous, arising from reporting lags, delay in verification and processing, communication delays, technological constraints, or regulatory requirements.

*3.2 Assumptions*

**Assumption 1 (Diffusion Dynamics).** The underlying economic state evolves as a continuous-time diffusion process, $dX_t = \mu(X_t, t)dt + \sigma(X_t, t)dW_t$, where $\mu(X_t, t)$ denotes the drift of the underlying state; $\sigma(X_t, t)$ the diffusion coefficient; and $W_t$ is the standard Brownian motion. The drift ($\mu$) and diffusion ($\sigma$) functions are twice continuously differentiable and satisfy the standard Lipschitz and linear-growth conditions ensuring a unique strong solution (Øksendal, 2003). The diffusion is nondegenerate: $\sigma^2(x, t) > 0$ for all $(x, t)$.

**Assumption 2 (Finite Second Moments).** The payoff-relevant state satisfies $\mathbb{E}[X_t^2] < \infty \ \forall t$.

**Assumption 3 (Rational Expectations).** Decision makers know the true stochastic law governing the underlying state and, under mean-squared error loss, form the optimal forecast: $\hat{X}_t = \mathbb{E}[X_t \mid \mathcal{F}_{t-\tau}]$.

**Assumption 4 (Accurate Observation).** The state is observed without measurement error at each observation date.

## 4. The Statistical Consequences of Information Latency

This section characterizes the statistical implications of the information structure introduced in Section 3, beginning with the local effect of latency on conditional variance before turning to its state- and timing-dependent extensions.

### *4.1 Forecasting under Delayed Information*

Consider a decision maker acting at time *t* whose information set is $\mathcal{F}_{t-\tau}$. Under Assumption 3, the optimal forecast of the contemporaneous state is $\hat{X}_t = \mathbb{E}[X_t \mid \mathcal{F}_{t-\tau}]$.

Define the forecast error as $\varepsilon_t = X_t - \hat{X}_t$. By the law of iterated expectations, $\mathbb{E}[\varepsilon_t \mid \mathcal{F}_{t-\tau}] = 0$: information latency generates no systematic forecast bias, since rational expectations fully incorporate all available information at the decision time.

Unbiasedness, however, does not imply perfect accuracy. Because the underlying state continues to evolve between $t - \tau$ and *t*, uncertainty remains about its contemporaneous value measured by $Var(X_t \mid \mathcal{F}_{t-\tau}) = \mathbb{E}[\varepsilon_t^2 \mid \mathcal{F}_{t-\tau}]$. For two latency levels $\tau_2 > \tau_1 \geq 0$, $\mathcal{F}_{t-\tau_2} \subseteq \mathcal{F}_{t-\tau_1}$. Because the information structure is nested, a decision maker with the more recent information set can always reproduce any decision available under the older information set. Less latency is therefore weakly preferred ex ante in every decision problem (Blackwell, 1953). This ordering is qualitative, however: it does not determine the magnitude of the value of freshness, nor does it imply that posterior uncertainty is ordered after every particular information realization. Sections 4.2 and 4.3 address these two questions.

### *4.2 Information Latency Theorem*

Diffusion dynamics imply $X_t = X_{t-\tau} + \int_{t-\tau}^{t} \mu(X_s, s)ds + \int_{t-\tau}^{t} \sigma(X_s, s)dW_s$. The first integral is the drift component of the state's evolution over the latency interval, while the second captures the cumulative stochastic innovations occurring between $t - \tau$ and *t*. Because these innovations occur after $t - \tau$ and therefore are not measurable with respect to $\mathcal{F}_{t-\tau}$, they generate residual uncertainty about the contemporaneous state even under rational expectations.

**Theorem 1 (Statistical Consequences of Information Latency).** Suppose Assumptions 1–4 hold and $\tau > 0$. Then: 1) the optimal forecast is conditionally unbiased, $\mathbb{E}[X_t \mid \mathcal{F}_{t-\tau}] = \hat{X}_t$; 2) conditional uncertainty is strictly positive, $Var(X_t \mid \mathcal{F}_{t-\tau}) > 0$; and, 3) for sufficiently small $\tau$, $Var(X_t \mid \mathcal{F}_{t-\tau}) = \sigma^2(X_{t-\tau}, t - \tau)\tau + O(\tau^2)$. Consequently, conditional uncertainty increases linearly with information latency to first order, with the local rate of increase determined by the diffusion variance of the underlying state.

***Proof.*** Provided in Appendix A.1.

*Interpretation*. Rational expectations eliminate predictable forecast errors but cannot eliminate uncertainty arising from stochastic innovations during the latency interval. This result is local, however. It establishes the direction and magnitude of the effect for short delays only without imposing a global monotonicity claim after every possible information realization—a distinction that becomes important in Section 4.3.

*Remark (Beyond Diffusions).* The linear local expansion is not confined to Itô diffusions. Let $\{X_t\}$ be a sufficiently regular time-homogeneous Markov process with infinitesimal generator $\mathcal{A}$, and $f(x) = x$, with $f, f^2 \in \mathcal{D}(\mathcal{A})$ (Ethier and Kurtz, 1986). Conditional on $X_{t-\tau} = x$, $\mathrm{Var}_x(X_\tau) = [\mathcal{A}(f^2)(x) - 2f(x)\mathcal{A}f(x)]\tau + o(\tau)$. For an Itô diffusion, the coefficient of $\tau$ equals $\sigma^2(x)$, recovering the leading-order term in Theorem 1. The $O(\tau^2)$ remainder in Theorem 1 follows from the stronger regularity available in the diffusion setting. Thus the diffusion model provides a tractable benchmark for a more general local property of continuous-time Markov processes.

**Example (Mean-Reverting State).** Consider the Ornstein–Uhlenbeck (OU) process $dX_t = \kappa(\theta - X_t)dt + \sigma dW_t$, where $\theta$ is long-run mean and $\kappa > 0$ is the speed of mean reversion. Conditional variance is $Var(X_t \mid \mathcal{F}_{t-\tau}) = \frac{\sigma^2}{2\kappa}(1 - \mathrm{e}^{-2\kappa\tau})$, which equals $\sigma^2\tau + O(\tau^2)$ for sufficiently small $\tau$, consistent with Theorem 1. Moreover,

$$\frac{\partial \mathrm{Var}(X_t \mid \mathcal{F}_{t-\tau})}{\partial \tau} = \sigma^2 \mathrm{e}^{-2\kappa\tau} > 0, \qquad \frac{\partial^2 \mathrm{Var}(X_t \mid \mathcal{F}_{t-\tau})}{\partial \tau^2} = -2\kappa\sigma^2 \mathrm{e}^{-2\kappa\tau} < 0.$$

Thus, under mean reversion, latency increases uncertainty but at a decreasing rate. As $\tau \to \infty$, this uncertainty converges to $\sigma^2/2\kappa$. Mean reversion therefore places a finite upper bound on both latency-induced variance and the associated risk compensation. In contrast, in the limiting case $\kappa \to 0$, the process reduces to Brownian motion and $Var(X_t \mid \mathcal{F}_{t-\tau}) = \sigma^2\tau$, which grows without bound as $\tau \to \infty$.

**Corollary 1 (Local Effect of Information Latency).** Under the assumption of Theorem 1, let $v(s, x;\ \tau) = Var(X_{s+\tau} \mid X_s = x)$. Holding *s* and *x* fixed,

$$\lim_{\tau \downarrow 0} \frac{v(s, x;\ \tau) - v(s, x;\ 0)}{\tau} = \sigma^2(x, s) > 0.$$

Hence, conditional uncertainty is locally increasing in information latency, with its initial rate determined by the state's instantaneous variance.

***Proof.*** Provided in Appendix A.2.

*4.3 Ex Ante Monotonicity and State-Conditional Latency Reversal*

Theorem 1's local result does not imply that uncertainty must continue to increase over the entire range of latency after every particular lagged-state observation. The distinction is between expected conditional uncertainty, which averages across possible observations, and state-conditional uncertainty following a specific observation.

**Theorem 2 (Ex Ante Monotonicity and State-Conditional Latency Reversal).** Let $X_t$ have a finite second moment. Then,

1) For any $t \geq \tau_2 > \tau_1 \geq 0$, $\mathbb{E}\left[\mathrm{Var}\left(X_t \mid \mathcal{F}_{t-\tau_1}\right)\right] \leq \mathbb{E}\left[\mathrm{Var}\left(X_t \mid \mathcal{F}_{t-\tau_2}\right)\right]$. Thus, expected conditional uncertainty is weakly increasing in latency, and
2) State-conditional uncertainty need not increase monotonically with latency.

Following Norris (1997), let $X_t \in \{0, 1\}$ follow a continuous-time Markov chain with generator

$$Q = \begin{pmatrix} -\lambda_{01} & \lambda_{01} \\ \lambda_{10} & -\lambda_{10} \end{pmatrix}, \qquad \lambda_{01}, \lambda_{10} > 0,$$

with jump rate $\lambda = \lambda_{01} + \lambda_{10}$. Let $\bar{p} = \lambda_{01}/\lambda$ be the stationary probability of state 1. Conditional on observing $X_{t-\tau} = x \in \{0, 1\}$, define $p_x(\tau) \equiv \mathbb{P}(X_t = 1 \mid X_{t-\tau} = x)$ and $v_x(\tau) \equiv \mathrm{Var}(X_t \mid X_{t-\tau} = x) = p_x(\tau)[1 - p_x(\tau)]$.

The transition probabilities are $p_0(\tau) = \bar{p}\left(1 - e^{-\lambda\tau}\right)$ and $p_1(\tau) = \bar{p} + (1 - \bar{p})e^{-\lambda\tau}$. If $\bar{p} > 1/2$, then, following an observation of state 0, $v_0(\tau)$ is strictly increasing in $0 < \tau < \tau_0^*$, strictly decreasing in $\tau > \tau_0^*$, and reaches its maximum value $1/4$ at

$$\tau_0^* = \frac{1}{\lambda}\log\left(\frac{2\bar{p}}{2\bar{p} - 1}\right).$$

If $\bar{p} < 1/2$, then, following an observation of state 1, $v_1(\tau)$ is strictly increasing in $0 < \tau < \tau_1^*$, strictly decreasing in $\tau > \tau_1^*$, and reaches its unique maximum value $1/4$ at

$$\tau_1^* = \frac{1}{\lambda}\log\left(\frac{2(1 - \bar{p})}{1 - 2\bar{p}}\right).$$

3) If the two-state process is stationary, then $\mathbb{E}[\mathrm{Var}(X_t \mid X_{t-\tau})] = \bar{p}(1 - \bar{p})\left(1 - e^{-2\lambda\tau}\right)$, which is strictly increasing in $\tau$.

***Proof.*** Provided in Appendix A.3.

*Interpretation.* States 0 and 1 are labels for two possible economic regimes—for example state 0 as financial distress and state 1 as normal operation. If $\bar{p} > 1/2$, state 1 is the more common long-run state and state 0 is relatively rare. Immediately after state 0 is observed, the current state remains relatively predictable. As latency increases, a transition to state 1 becomes more plausible, and uncertainty peaks when the two states are equally likely. At longer latencies, the rare-state observation loses relevance, beliefs converge toward the stationary distribution, and conditional variance declines toward its stationary value. Then following a rare-state observation, moderately stale information can generate greater state-conditional uncertainty than very stale information—a non-monotonic relationship we call a state-conditional reversal.

*Remark (Blackwell Ordering).* The reversal does not imply that stale information is generally more valuable, nor does it contradict the Blackwell ordering (Blackwell, 1953). A more recent information set contains the older one and is therefore weakly more valuable ex ante. Consistent with this ordering, Theorem 2 shows that expected conditional uncertainty increases with latency. The reversal concerns only uncertainty conditional on one particular lagged-state observation.

*4.4 Random Latency and Update Regularity*

In many applications, such as reporting schedules and periodic disclosures, information does not arrive at a single fixed interval. This section allows latency to be random while preserving exogenous information timing and exact observation. In doing so, the analysis distinguishes two conceptually different sources of variability: dispersion in the information age experienced at decision times and dispersion in the update intervals that generate that age.

4.4.1 Dispersion in Realized Latency

Let $A \geq 0$ denote the realized age of the most recent observation at the decision time. Assume $A$ is observed by the decision maker and is independent of subsequent state innovations. Conditional on $A = a$, the decision maker observes $\mathcal{F}_{t-a}$. For the OU process, conditional variance at realized latency $a$ is

$$V(a) \equiv \mathrm{Var}(X_t \mid \mathcal{F}_{t-a}) = \frac{\sigma^2}{2\kappa}(1 - e^{-2\kappa a}),$$

which is increasing and strictly concave: $V'(a) = \sigma^2 e^{-2\kappa a} > 0$, and $V''(a) = -2\kappa\sigma^2 e^{-2\kappa a} < 0$.

**Corollary 2 (Dispersion in Realized Latency).** Let $A_1$ and $A_2$ be nonnegative random latencies with the same finite mean. If $A_2$ is a mean-preserving spread of $A_1$, then, for the OU process, $\mathbb{E}[V(A_2)] \leq \mathbb{E}[V(A_1)]$, strictly for a nondegenerate mean-preserving spread. Because $\mathrm{V}(a)$ is concave, Jensen's inequality gives $\mathbb{E}[V(A)] \leq V(\mathbb{E}[A]) = \frac{\sigma^2}{2\kappa}\left(1 - e^{-2\kappa\mathbb{E}[A]}\right)$, with strict inequality whenever $A$ is nondegenerate.

***Proof.*** Provided in Appendix A.4.

*Interpretation.* Because conditional variance is concave in information age, holding mean realized latency fixed, greater dispersion lowers average conditional variance under mean reversion by Jensen's inequality. Intuitively, fresh observations produce large reductions in uncertainty, whereas additional staleness at already high latency adds progressively less uncertainty as variance approaches its stationary upper bound. This does not mean that greater realized latency reduces uncertainty. Nor does it imply that irregular information arrival is preferable overall. Greater dispersion may worsen tail losses, worst-case risk, operational costs, or other outcomes not captured by average conditional variance.

In the Brownian-motion limit, conditional variance is linear in latency: $V(a) = \sigma^2 a$. Average conditional variance therefore depends only on mean latency, not its dispersion: $\mathbb{E}[V(A)] = \sigma^2\mathbb{E}[A]$. The dispersion result is therefore driven by the concavity generated by mean reversion, rather than by randomness in latency itself.

4.4.2 Renewal Timing and Update Regularity

Let $0 = S_0 < S_1 < S_2 < \cdots$ denote successive update times, with inter-update intervals $T_n = S_n - S_{n-1}$. Assume that $\{T_n\}_{n\geq 1}$ are independent and identically distributed positive random variables, exogenously generated and satisfying $\mathbb{E}[T_n] < \infty, \mathbb{E}[T_n^2] < \infty$. The resulting sequence of update times forms a renewal process: after each update, the waiting time until the next update is drawn anew from the same distribution, independent of any agent's choice.

Let $T$ denote a generic inter-update interval having the common distribution of the $T_n$. At time $t$, define the number of completed updates by $N(t) = \max\{n: S_n \leq t\}$ and the age of the most recent update by $A(t) = t - S_{N(t)}$. Suppose the decision time is independent of the renewal process and samples it from its stationary long-run distribution.

**Theorem 3 (Update Regularity and Average Latency-Induced Variance).** Under the renewal update process:

1) Long-run expected information age is

$$\mathbb{E}[A] = \frac{\mathbb{E}[T^2]}{2\mathbb{E}[T]} = \frac{\mathbb{E}[T]}{2} + \frac{\mathrm{Var}(T)}{2\mathbb{E}[T]};$$

2) For a Brownian state with variance rate $\sigma^2$, long-run average latency-induced variance is

$$\overline{V}_B = \frac{\sigma^2}{2}\left[\mathbb{E}[T] + \frac{\mathrm{Var}(T)}{\mathbb{E}[T]}\right];$$

3) For an OU state, long-run average latency-induced variance is

$$\overline{V}_{OU} = \frac{\sigma^2}{2\kappa}\left[1 - \frac{1 - \mathbb{E}(e^{-2\kappa T})}{2\kappa\mathbb{E}[T]}\right].$$

Among renewal processes with a given mean inter-update interval, deterministic updating minimizes long-run average latency-induced variance. More generally, if $T_2$ is a mean-preserving spread of $T_1$, then $\overline{V}(T_2) \geq \overline{V}(T_1)$ for both Brownian and OU dynamics, with strict inequality for a nondegenerate mean-preserving spread.

***Proof.*** Provided in Appendix A.5.

*Interpretation.* A randomly selected decision time is more likely to fall within a long update interval than a short one—the renewal-theoretic inspection effect (Feller, 1971). Thus, two systems sharing the same mean inter-update interval can still differ in decision-time information ages and latency cost.

Consider two systems updating once every 30 days on average. With deterministic 30-day intervals, information age rises uniformly from 0 to 30 days, averaging 15 days. With Poisson updating, *T* is exponentially distributed with mean 30 days and mean realized information age is also $\mathbb{E}[A] = \frac{30}{2} + \frac{30^2}{2(30)} = 30$ days.

The irregular system has the same average update frequency but twice the average realized latency. Under Brownian dynamics, it therefore generates twice the average latency-induced variance. Reporting standards based only on average update frequency can therefore conceal economically important differences in information quality: arrival regularity matters even when the long-run update rate is unchanged.

Corollary 2 and Theorem 3 appear to reach opposite conclusions. However, Corollary 2 compares distributions of realized latency $A$, holding $\mathbb{E}[A]$ fixed. In contrast, Theorem 3 compares distributions of the inter-arrival interval $T$, holding $\mathbb{E}[T]$ fixed. Thus, greater variability in experienced latency at a fixed mean age can reduce average uncertainty, while greater variability in information-arrival intervals at a fixed mean interval increases it.

## 5. Economic Consequences of Information Latency

This section studies how decision makers respond to the irreducible uncertainty generated by delayed information. We first present a general decision-theoretic framework that applies to any optimization problem under uncertainty and then derive a closed-form Latency Premium Equation under Constant Absolute Risk Aversion (CARA) preferences.

### *5.1 Decision Making under Information Latency*

Consider a decision maker choosing an action $a \in \mathcal{A} \subseteq \mathbb{R}$ at time $t$ and let $u : \mathbb{R} \to \mathbb{R}$ be a strictly increasing, concave von Neumann–Morgenstern utility function. The decision maker's realized payoff is $\pi = \pi(a, X_t)$. We assume $\pi(\cdot, x)$ is continuous and concave in $a$ for each $x$ and $\mathcal{A}$ is compact, so a maximizer exists. Because the current state is not directly observed, the decision maker chooses $a \in \mathcal{A}$, using only information available at $t - \tau$ to maximize,

$$\max_{a \in \mathcal{A}} E\,[u(\pi(a, X_t)) \mid \mathcal{F}_{t-\tau}].$$

This formulation encompasses a broad class of economic problems: the interest rate or lending decision in credit markets, the premium or contract offered in insurance, portfolio allocation or asset valuation in financial markets, inventory or production decisions in supply-chain management, and the policy interventions that rely on delayed economic indicators.

Latency does not alter the form of the optimization problem; it changes the conditional distribution of the payoff-relevant state. As established in Section 4.1, for $\tau_2 > \tau_1$, the decision maker with latency $\tau_1$ can always reproduce any strategy available under $\tau_2$. Optimized expected utility is therefore weakly higher with the more recent information set.

### *5.2 The Economic Cost of Conditional Uncertainty*

Fix an action $a$ and let $\hat{\pi} \equiv \mathbb{E}[\pi(a, X_t) \mid \mathcal{F}_{t-\tau}]$ denote conditional expected payoff. A second-order Taylor expansion of $u(\pi)$ around $\hat{\pi}$ gives

$$\mathbb{E}[u(\pi) \mid \mathcal{F}_{t-\tau}] \approx u(\hat{\pi}) + \frac{1}{2} u''(\hat{\pi}) \mathrm{Var}(\pi \mid \mathcal{F}_{t-\tau}).$$

Because $u''(\hat{\pi}) < 0$,

$$\frac{\partial\, \mathbb{E}[u(\pi) \mid \mathcal{F}_{t-\tau}]}{\partial\, \mathrm{Var}(\pi \mid \mathcal{F}_{t-\tau})} \approx \frac{1}{2} u''(\hat{\pi}) < 0.$$

To second order, holding the conditional mean fixed, greater conditional variance lowers expected utility. Combining this approximation with Theorem 1 yields the local variance channel: $\mathrm{Var}(X_t \mid \mathcal{F}_{t-\tau}) = \sigma^2\tau + O(\tau^2)$. Accordingly, for sufficiently small latency, an increase in latency reduces expected utility through the first-order increase in unresolved state uncertainty.

*5.3 A Constant Absolute Risk Aversion (CARA) Benchmark*

To obtain a closed-form expression for the latency premium, suppose preferences exhibit CARA, $u(\pi) = -e^{-\gamma\pi},\ \gamma > 0$, where γ denotes the coefficient of absolute risk aversion.

**Assumption 5 (Conditional Normality).** Conditional on $\mathcal{F}_{t-\tau}$, the payoff-relevant state, $X_t$, is approximately normally distributed: $X_t \mid \mathcal{F}_{t-\tau} \sim N\big(\hat{X}_t, Var(X_t \mid \mathcal{F}_{t-\tau})\big)$, with $\hat{X}_t$ and $Var(X_t \mid \mathcal{F}_{t-\tau})$ as characterized in Theorem 1. This is stronger than what is required for Theorem 1, which characterizes the first two conditional moments. It holds exactly in the OU special case of Assumption 1 and serves as a tractable local approximation.

For simplicity, we also suppose the payoff is linear in the underlying state, $\pi = P + \beta X_t$, where $P$ is the state-independent component of the payoff and $\beta$ measures the sensitivity of the payoff to the underlying state.

**Proposition 1 (Latency Premium).** Under Assumptions 1–5 and CARA preferences with linear state exposure, the certainty equivalent (CE) takes the standard mean-variance form,

$$CE(\tau) = P + \beta\hat{X}_t - \frac{\gamma\beta^2}{2}\mathrm{Var}(X_t \mid \mathcal{F}_{t-\tau}).$$

The compensation required for bearing latency-induced conditional risk is therefore

$$\Pi(\tau) = \frac{\gamma\beta^2}{2}\mathrm{Var}(X_t \mid \mathcal{F}_{t-\tau}).$$

***Proof.*** Provided in Appendix A.6.

Under the OU benchmark, the premium is globally increasing but concave in latency, converging to

$$\lim_{\tau\to\infty}\Pi(\tau) = \frac{\gamma\beta^2\sigma^2}{4\kappa}.$$

Mean reversion therefore bounds both conditional variance and the associated compensation.

**Corollary 3 (Sensitivity of the Latency Premium).** Under Proposition 1 and Theorem 1's variance expansion, the latency premium $\Pi(\tau)$ admits the local marginal sensitivities:

$$\frac{\partial \Pi}{\partial \tau} = \frac{\gamma\beta^2}{2}\sigma^2 + O(\tau) \text{ and } \frac{\partial \Pi}{\partial \sigma^2} \approx \frac{\gamma\beta^2}{2}\tau,$$

each of which is strictly positive for non-zero exposure and sufficiently small $\tau > 0$.

***Proof.*** Provided in Appendix A.7.

*Interpretation.* For sufficiently short delays, the risk premium increases with latency. The premium also increases with the state's instantaneous variance: when the underlying state evolves more rapidly, a given delay generates more uncertainty. Risk aversion and payoff exposure enter as scaling factors, magnifying both marginal effects proportionally.

**Corollary 4 (Value of Timely Information).** Under Proposition 1, and holding the conditional forecast fixed, certainty-equivalent value satisfies $\partial CE(\tau)/\partial\tau < 0$ for sufficiently small $\tau$. In the OU benchmark, this inequality holds for every finite $\tau > 0$.

***Proof.*** Provided in Appendix A.8.

*5.4 State-Conditional Latency Premium*

Sections 4.3 and 4.4 showed how conditional uncertainty depends on the last observed state and on information timing. This section explores the corresponding implications for risk compensation.

Return to the two-state Markov process of Theorem 2 and suppose $\pi = P + \beta X_t$, where $\beta > 0$. Conditional on $p = \mathbb{P}(X_t = 1 \mid X_{t-\tau} = x)$, expected utility is $\mathbb{E}[u(\pi) \mid p] = -e^{-\gamma P}\left(1 - p + pe^{-\gamma\beta}\right)$. The corresponding certainty equivalent is $CE(p) = P - \frac{1}{\gamma}\log\left(1 - p + pe^{-\gamma\beta}\right)$, and the exact risk premium relative to conditional expected payoff is

$$\Pi(p) = \beta p + \frac{1}{\gamma}\log\left(1 - p + pe^{-\gamma\beta}\right).$$

**Proposition 2 (State-Conditional Latency Premium).** The binary-state premium $\Pi(p)$ is strictly concave in $p$, satisfies $\Pi(0) = \Pi(1) = 0$, and has a unique maximum at

$$p^{\dagger} = \frac{1}{1 - e^{-\gamma\beta}} - \frac{1}{\gamma\beta}.$$

For small $\gamma\beta$,

$$p^{\dagger} = \frac{1}{2} + \frac{\gamma\beta}{12} + O((\gamma\beta)^3).$$

Consequently, if the limiting probability of state 1 is $\bar{p}$, then following an observation of state 0, the latency premium is hump-shaped in $\tau$ whenever $\bar{p} > p^{\dagger}$. Following an observation of state 1, it is hump-shaped whenever $\bar{p} < p^{\dagger}$.

***Proof.*** Provided in Appendix A.9.

*Interpretation.* Proposition 2 shows how the state-conditional non-monotonicity in Theorem 2 carries over to risk compensation. Because the exact binary-state premium is maximized at an interior posterior $p^{\dagger}$, required risk compensation can initially increase and subsequently decrease with latency whenever the posterior path induced by increasing latency crosses $p^{\dagger}$.

Such non-monotonicity is especially relevant following an observation of a relatively unlikely state, because increasing latency moves beliefs away from the observed state and back toward the stationary distribution.

This state-conditional reversal does not imply that stale information is ex ante preferable to fresh information. It is an ex post effect conditional on a particular previous observation, not a reversal of the ex ante value of information.

The preceding result concerns the exact premium in the binary-state model. A separate implication arises when we return to the CARA-normal benchmark and allow latency itself to be generated by the irregular renewal timing analyzed in Section 4.4.2. Define the long-run average of premiums conditional on realized and observed age $A$ by

$$\overline{\Pi} \equiv \mathbb{E}[\Pi(A)] = \frac{\gamma\beta^2}{2}\mathbb{E}[V(A)].$$

**Corollary 5 (Average Premium under Random Latency).** Under Proposition 1:

1) Holding mean realized latency fixed, a mean-preserving spread in age $A$ lowers the average conditional latency premium under OU dynamics;
2) Holding the mean inter-update interval fixed, a mean-preserving spread in renewal intervals $T$ raises the average conditional latency premium under both Brownian and OU dynamics.

For Brownian dynamics,

$$\overline{\Pi}_B = \frac{\gamma\beta^2\sigma^2}{4}\left[\mathbb{E}[T] + \frac{\mathrm{Var}(T)}{\mathbb{E}[T]}\right].$$

For OU dynamics,

$$\overline{\Pi}_{OU} = \frac{\gamma\beta^2\sigma^2}{4\kappa}\left[1 - \frac{1 - \mathbb{E}(e^{-2\kappa T})}{2\kappa\mathbb{E}[T]}\right].$$

***Proof.*** Provided in Appendix A.10.

*Interpretation.* Corollary 5 translates the random-latency results of Corollary 2 and Theorem 3 directly into economic compensation. Under the CARA-normal benchmark, the corresponding effects on average conditional variance carry over proportionally to the average latency premium. Thus, dispersion in realized age and dispersion in renewal intervals continue to have opposite effects when their respective means are held fixed, as explained in Section 4.4.2.

Importantly, these average premiums condition on realized and observed age. If compensation is determined before *A* is realized, the relevant ex ante certainty equivalent is computed from the full mixture distribution over ages. Because certainty equivalents are nonlinear, the resulting ex ante risk premium generally differs from $\mathbb{E}[\Pi(A)]$.

*5.5 Bertrand Competition with Heterogeneous Latency*

The baseline framework characterizes a single decision maker in isolation. In decentralized markets—such as consumer lending or insurance underwriting— multiple intermediaries compete under different latency constraints.

Consider *N* risk-averse intermediaries with CARA preferences, indexed by $i \in \{1, \dots, N\}$, with risk aversion $\gamma_i$ and latency $\tau_i$. Each intermediary offers a contract with payoff $\pi_i = P_i + \beta X_t$ tied to the client's current latent state $X_t$. Intermediaries engage in Bertrand price competition to provide coverage or credit up to reservation certainty equivalent *C*.

To isolate differences in latency-induced risk, suppose all intermediaries share a common conditional forecast $\hat{X}_t$ and differ only in their risk aversion and latency-induced conditional variance. This assumption isolates the risk-compensation channel by holding differences in conditional means fixed; in general, heterogeneous latency may affect both the conditional forecast and conditional variance. Define intermediary *i*'s reservation price as

$$r_i \equiv C - \beta\hat{X}_t + \Pi_i(\tau_i), \quad \text{where } \Pi_i(\tau_i) \equiv \frac{\gamma_i\beta^2}{2}\mathrm{Var}\left(X_t \mid \mathcal{F}_{t-\tau_i}\right).$$

Order the reservation prices so that $r_1 < r_2 \leq \ldots \leq r_N$, and let $i^*$ denote the unique intermediary with reservation price $r_1$. Each intermediary posts a price, and the client selects the lowest-priced otherwise identical contract.

**Proposition 3 (Bertrand Pricing with Heterogeneous Latency Costs).** Under Bertrand competition, intermediary $i^*$ with the lowest risk-adjusted reservation cost serves the market, while the limiting price is determined by the second-lowest reservation cost:

$$P^* = r_{(2)} = C - \beta \hat{X}_t + \Pi_{(2)},$$

where $\Pi_{(2)} \equiv \min_{k \neq i^*} \Pi_k$. The winning intermediary earns certainty-equivalent rent $P^* - r_{(1)} = \Pi_{(2)} - \Pi_{(1)} > 0$. If ties at $r_{(2)}$ are resolved in favor of the lowest-cost intermediary, the limit price is attained. Otherwise, the winner posts $r_{(2)} - \varepsilon$, with $\varepsilon \downarrow 0$.

***Proof.*** Provided in Appendix A.11.

*Remark* (*Ranking Intermediaries*). Under globally monotonic benchmarks such as the OU model, lower latency implies a lower risk-adjusted cost, all else equal. Under the state-conditional reversal in Theorem 2, however, latency alone need not determine this ranking after every observed state. The relevant object is the resulting premium. Likewise, with random updates, average update frequency alone is insufficient because update regularity affects their long-run latency costs.

**Corollary 6 (Contestability and the Information-Efficiency Frontier).** If competition at the technological frontier intensifies so that $\Pi_{(2)} - \Pi_{(1)} \to 0$, then the winner's rent vanishes. If, in addition, $\Pi_{(2)} \to 0$ and the common forecast converges to the contemporaneous state, then $P^* \to C - \beta X_t$. The price approaches the frictionless benchmark only when at least two competitors approach the information-efficiency frontier or entry drives the second-lowest risk-adjusted cost toward zero.

***Proof.*** Immediate from Proposition 3, taking $\Pi_{(2)} - \Pi_{(1)} \to 0$.

*Interpretation.* Better information lowers an intermediary's risk-bearing cost, but competition determines how much of that saving reaches clients. A sole frontier intermediary can retain an informational rent because its price is disciplined by its nearest rival's cost. As rivals catch up, this rent disappears. Eliminating rents does not, however, eliminate the cost of delayed information: prices approach the frictionless benchmark only when competing suppliers also approach zero latency cost and forecasts approach the current state.

*5.6 Latency as a Second-Best Commitment Device*

Although fresher information lowers the cost of bearing uncertainty in the benchmark model, it can reduce welfare by undermining risk-sharing opportunities, as Hirshleifer (1971) shows. In insurance markets, Hendel and Lizzeri (2003) show how long-term contracts can protect policyholders against reclassification risk. We examine whether a deliberately chosen observation lag can provide a related form of protection by limiting the frequency with which new information can trigger reclassification. In this sense, latency can serve as a second-best commitment device when direct contractual restrictions on reclassification are unavailable.

Consider a contract initiated at date 0 and renewed at later date $T > 0$. At initiation, the policyholder's state $X_0$ is commonly observed and lies below the reclassification threshold, $\bar{X}$. For tractability, suppose that the state follows the Brownian benchmark $X_s = X_0 + \sigma W_s$.

At renewal, an insurer with latency $\tau \in [0, \mathrm{T}]$ observes $X_{T-\tau}$. If this observation exceeds $\bar{X}$, the policyholder is reclassified or denied renewal and incurs an incremental certainty-equivalent loss $(L > 0)$, representing the welfare cost of less favorable coverage terms or loss of coverage. Otherwise, coverage continues and the policyholder pays a contract price that includes the latency premium derived in Section 5.3.

Let $\Delta \equiv \bar{X} - X_0 > 0$. For $0 \le \tau < T$, the probability of reclassification is

$$p(\tau) \equiv \mathbb{P}(X_{T-\tau} > \bar{X} \mid X_0) = 1 - \Phi\left(\frac{\Delta}{\sigma\sqrt{T-\tau}}\right),$$

where $\Phi$ is the standard normal distribution function. At $\tau = \mathrm{T}$, the insurer observes the initial state $X_0$, which lies below the threshold by assumption. Hence $p(T) = 0$, which is also the continuous limit: $\lim_{\tau \to T} p(\tau) = 0$.

A longer lag shortens the interval between the initial observation and the observation used at renewal. Under the Brownian benchmark, therefore $p(\tau)$ is decreasing in $\tau$. Under the CARA-normal benchmark, the latency premium is

$$\Pi(\tau) = \frac{\gamma\beta^2\sigma^2}{2}\tau.$$

Assuming this premium is fully passed through to the policyholder, define the expected incremental cost relative to the actuarially fair base price as $J(\tau) = [1 - p(\tau)]\Pi(\tau) + p(\tau)L$. The first term is the expected cost of latency when coverage continues; the second is the expected loss from reclassification. Latency thus trades a higher cost of continued coverage against a lower probability of reclassification.

**Proposition 4 (Latency as a Second-Best Commitment Device).** Let $p_0 = p(0)$. If

$$L > \frac{\gamma\beta^2\sigma^2}{2}\frac{(1-p_0)}{|p'(0)|},$$

then zero latency does not minimize $J(\tau)$, and at least one optimal latency satisfies $\tau^* \in (0, T)$. Moreover,

$$|p'(0)| = \frac{\Delta}{2\sigma T^{3/2}}\phi\left(\frac{\Delta}{\sigma\sqrt{T}}\right),$$

where $\phi$ denotes the standard normal density. Equivalently, the sufficient condition is

$$L > \frac{\gamma\beta^2\sigma^3 T^{3/2}(1-p_0)}{\Delta\,\phi\left(\Delta/(\sigma\sqrt{T})\right)}.$$

***Proof.*** Provided in Appendix A.12.

*Interpretation.* Latency operates through two opposing channels. It raises the premium required for unresolved current-state uncertainty, but reduces the probability of reclassification based on adverse information. When reclassification is sufficiently costly, a small increase in latency from zero lowers the policyholder's expected incremental cost. As latency approaches renewal horizon, its marginal protection against reclassification disappears while its marginal uncertainty cost remains positive, producing an interior optimum.

This benefit is second-best. If an enforceable contract can prevent reclassification while allowing the insurer to observe current information, fresh information combined with a restriction on its use weakly dominates deliberate informational delay. Latency can therefore provide partial protection when direct contractual commitment is unavailable. The result distinguishes the value of observing information from the consequences of allowing it to affect contract terms.

## 6. Applications

The applications below illustrate several economic consequences of information latency across credit, insurance, financial markets, and operational decisions.

*6.1 Consumer Credit*

Lenders often rely on historical financial records even as borrowers' income, employment, and indebtedness change. Credit-account update is typically reported in monthly batches (Consumer Financial Protection Bureau, 2012). Berg et al. (2020) document the predictive value of digital footprints beyond credit-bureau scores, although they do not isolate the effect of information freshness. In our framework, older borrower information raises the compensation required for unresolved repayment risk, particularly when underlying financial conditions are volatile.

A distinctive implication arises after an adverse signal. Suppose delinquency is rare relative to normal repayment. Immediately after delinquency is observed, the borrower's state may be relatively predictable. As that observation ages, uncertainty can first increase as both recovery and continued delinquency become plausible, and then decline as beliefs revert toward the predominantly normal stationary distribution. The uncertainty component of the lending premium can therefore be hump-shaped in the age of an adverse report, even though the total loan spread need not be, since expected losses and other pricing components may evolve simultaneously.

The framework therefore isolates an information-timing channel distinct from the informational advantages associated with lending relationships and geographic proximity studied by Sharpe (1990) and Agarwal and Hauswald (2010). It also implies that improvements in reporting frequency should matter most for borrowers whose financial states are more volatile.

*6.2 Insurance and Remote Monitoring*

Insurance underwriting relies on observations of evolving health, driving, property, or other conditions. Fresher observations can reduce unresolved uncertainty and the associated latency premium. Fresher information can also have a different contractual effect: when new information triggers costly reclassification, positive latency can protect policyholders when direct contractual commitment is unavailable. This second-best role of latency complements the protection against reclassification risk provided by long-term contracts (Hendel and Lizzeri, 2003).

A separate issue concerns not how recent information is at a given moment, but how regularly new information arrives. Two information systems can have the same average update frequency while generating very different patterns of realized information age. For example, Landsat 8 and 9 jointly provide an eight-day revisit cycle, but cloud cover can prevent usable optical observations (U.S. Geological Survey, 2022, 2023). As a result, the economically relevant information age for crop insurers or environmental-risk assessors is measured from the last usable image rather than from the nominal satellite revisit interval. Effective latency can therefore be much longer and more variable than the nominal schedule suggests.

This distinction matters because irregular information arrival can increase the expected cost of delayed information even when average update frequency is unchanged. Long gaps receive disproportionate weight in the realized age distribution, making decision makers more likely to encounter stale information. In Landsat applications, however, observation failures may not be fully exogenous. Cloud-related gaps may themselves be correlated with environmental conditions. The renewal benchmark should therefore be interpreted as applying most directly when update timing is conditionally independent of the underlying state; otherwise, the timing of missing observations can itself be informative.

*6.3 Financial Markets*

Corporate disclosures describe firm conditions at a particular reporting date, but those conditions may change before investors act. The framework therefore complements research on disclosure dissemination and market outcomes (Bushee et al., 2010; Goldstein, Yang, and Zuo, 2023) by focusing on the uncertainty that accumulates as otherwise accurate disclosures become dated. The effect should be stronger for firms whose underlying fundamentals evolve more rapidly.

Latency-induced uncertainty may also appear in bid-ask spreads or analyst forecast dispersion, although neither is a direct measure of the latency premium. Spreads reflect trading frictions and information differences across market participants, while forecast dispersion reflects heterogeneity in beliefs and forecasting behavior.

Corporate bond markets provide a more direct setting for the risk-compensation channel. Reported leverage, liquidity, and other balance-sheet measures may accurately describe a firm at the reporting date yet become less informative about current conditions as the disclosure ages. The resulting uncertainty can contribute to the compensation investors require for holding corporate debt and hence to observed credit spreads, holding expected default losses and other spread determinants fixed.

*6.4 Supply-Chain and Operational Decisions*

Production and logistics decisions often rely on outdated observations of demand, inventories, transportation conditions, and production capacity. Temporary road closures, port disruptions, and production outages illustrate the state-conditional effect especially clearly. Immediately after a disruption is observed, continued disruption may remain relatively predictable. As that information ages, recovery becomes increasingly plausible, so uncertainty can rise before eventually declining as beliefs move toward a long-run distribution dominated by normal operation. Moderately stale information about a disruption can therefore generate more uncertainty than substantially older information.

The regularity of monitoring also matters. Two monitoring systems can provide the same average number of updates while generating very different distributions of realized information age. More irregular update schedules create a greater chance of long information gaps, which can raise the average cost of stale information even when mean update frequency is unchanged. For supply-chain managers, this implies that the reliability and regularity of monitoring may matter independently of its average frequency.

## 7. Conclusion and Policy Implications

Information used in economic decisions is often accurate but outdated. This paper introduces information latency—the elapsed time between a decision and the most recent available observation of an evolving payoff-relevant state—as a distinct source of imperfect information. Delayed observation generates conditional uncertainty even when information is exact and decision makers understand the state's stochastic evolution. For short delays, conditional variance grows approximately linearly with latency, at a rate determined by the state's instantaneous variance. This local result extends beyond diffusions to sufficiently regular continuous-time Markov processes.

Latency alone, however, does not fully characterize information quality. Following a rare-state observation, conditional uncertainty and the associated risk premium can peak at an intermediate latency before declining as beliefs revert toward their long-run distribution. Information timing matters independently as well. Greater dispersion in realized age can reduce average conditional variance under mean-reverting dynamics, whereas greater irregularity in renewal intervals raises average information age and conditional variance when mean update frequency is held fixed. Information quality therefore depends on the interaction among information age, the observed state, and the process governing observation arrivals.

These statistical effects have direct economic consequences. Under the CARA-normal benchmark, the latency-induced risk premium is determined by conditional variance, risk aversion, and squared payoff exposure. Under Bertrand competition, the intermediary with the lowest risk-adjusted reservation cost serves the market, with the limiting price determined by the second-lowest reservation cost.

The results also have implications for the design of information systems and reporting standards. Policies that improve information infrastructure should be evaluated not only by the average frequency of updates but also by the delays they reduce, the regularity with which usable observations arrive, and the dynamics of the underlying state. Open-banking systems illustrate this point: more timely access to transaction data can reduce uncertainty about a consumer's current financial condition. More generally, two information systems with the same average update frequency can generate different economic costs if one produces more irregular or longer episodes of stale information.

A separate policy implication concerns the distinction between information freshness and information use. Fresher information generally improves decision value ex ante, but newly observed information may also trigger costly reclassification—for example, moving a borrower or policyholder into a less favorable risk category. When direct contractual commitment is unavailable, positive latency can provide second-best protection by limiting how frequently new information can be used to revise contract terms. When enforceable restrictions on information use are available, however, fresh information combined with those restrictions weakly dominates deliberate delay. The U.S. Card Act (2009) provides an illustration of this distinction: updated information can remain available even when its use for repricing existing contractual obligations is restricted. Similar considerations arise with open banking, where improved access to timely data may enhance risk assessment while also facilitating adverse contractual responses to newly observed financial distress.

Information policy should therefore distinguish three separate margins: how recent the available information is, how regularly new information arrives, and how newly acquired information may be used. Changes along these margins need not have the same welfare consequences. In particular, policies that increase data freshness need not imply unrestricted use of that information for reclassification or repricing.

Future work can extend the framework in several directions. One is to endogenize update frequency and renewal timing through costly information acquisition, especially when observation timing itself depends on the underlying state. A second is to study strategic investment in latency reduction when competing intermediaries choose both information technology and contractual terms. A third is to examine environments with multiple information sources that differ in age, precision, and sampling schedules, where the decision maker must optimally combine signals of different freshness. The framework also suggests empirical tests that separate the effect of information age from signal precision, expected losses, and other determinants of risk compensation. The central implication is that information quality is jointly determined by freshness, the observed state, and the timing process that generates observations. Two decision makers—or two information systems—with equally accurate data and the same average update frequency may therefore face materially different uncertainty and economic costs.

## Appendix

*A.1 Proof of Theorem 1.*

Unbiasedness: By Assumption 3, $\hat{X}_t = E[X_t \mid \mathcal{F}_{t-\tau}]$. Since $\hat{X}_t$ is $\mathcal{F}_{t-\tau}$ measurable, $E[\varepsilon_t \mid \mathcal{F}_{t-\tau}] = E[X_t \mid \mathcal{F}_{t-\tau}] - \hat{X}_t = 0$, which establishes conditional unbiasedness.

*Local Variance Expansion.* For sufficiently small $\tau$, Assumption 1's smoothness conditions imply the conditional first moment of the diffusion satisfies

$$E[X_t \mid \mathcal{F}_{t-\tau}] = X_{t-\tau} + \mu\,(X_{t-\tau}, t-\tau)\,\tau + O(\tau^2).$$

Next, applying Itô's formula to $X_s^2$ (Assumption 1's differentiability conditions),

$$dX_s^2 = [2X_s\,\mu(X_s, s) + \sigma^2(X_s, s)]ds + 2X_s\,\sigma(X_s, s)dW_s.$$

Integrating over $[t-\tau\,,\,t]$ and conditioning on $\mathcal{F}_{t-\tau}$, the Itô integral has conditional expectation zero, well defined by the finite-second-moment condition in Assumption 2. Therefore,

$$E[X_t^2 \mid \mathcal{F}_{t-\tau}] = X_{t-\tau}^2 + [2X_{t-\tau}\,\mu(X_{t-\tau}, t-\tau) + \sigma^2(X_{t-\tau}, t-\tau)]\,\tau + O(\tau^2).$$

By definition, $Var(X_t \mid \mathcal{F}_{t-\tau}) = \mathrm{E}[X_t^2 \mid \mathcal{F}_{t-\tau}] - (\mathrm{E}[X_t \mid \mathcal{F}_{t-\tau}])^2$, which is finite and well defined by Assumption 2. Using the first-moment expansion,

$$(\mathrm{E}[X_t \mid \mathcal{F}_{t-\tau}])^2 = X_{t-\tau}^2 + 2X_{t-\tau}\,\mu\,(X_{t-\tau}, t-\tau)\tau + O(\tau^2).$$

Subtracting yields, $Var(X_t \mid \mathcal{F}_{t-\tau}) = \sigma^2(X_{t-\tau}, t-\tau)\,\tau + O(\tau^2)$.

*Strict Positivity.* Because the diffusion is nondegenerate (Assumption 1), stochastic innovations occur over every positive latency interval, implying $Var(X_t \mid \mathcal{F}_{t-\tau}) > 0$ for $\tau > 0$. □

*A.2 Proof of Corollary 1.*

By Theorem 1, $v(s, x;\ \tau) = \sigma^2(x, s)\tau + O(\tau^2)$. Since $v(s, x;\ 0) = 0$,

$$\frac{v(s, x;\ \tau) - v(s, x;\ 0)}{\tau} = \sigma^2(x, s) + O(\tau).$$

Taking the limit as $\tau \downarrow 0$ gives the result, where the strict inequality follows from the nondegeneracy condition in Assumption 1. □

*A.3 Proof of Theorem 2.*

Let $\mathcal{G} = \mathcal{F}_{t-\tau_2}$, $\mathcal{H} = \mathcal{F}_{t-\tau_1}$, where $\tau_2 > \tau_1$, so $\mathcal{G} \subseteq \mathcal{H}$. Conditional variance decomposition gives $\mathbb{E}[\text{Var}(X_t \mid \mathcal{G})] = \mathbb{E}[\text{Var}(X_t \mid \mathcal{H})] + \mathbb{E}[\text{Var}(\mathbb{E}[X_t \mid \mathcal{H}] \mid \mathcal{G})]$. The final term is nonnegative, proving ex ante monotonicity. For the two-state chain, standard transition probabilities give $p_0(\tau) = \bar{p}(1 - e^{-\lambda\tau})$, and $p_1(\tau) = \bar{p} + (1 - \bar{p})e^{-\lambda\tau}$.

Because $X_t$ is conditionally Bernoulli, $v_x(\tau) = p_x(\tau)[1 - p_x(\tau)]$ and $v_x'(\tau) = p_x'(\tau)[1 - 2p_x(\tau)]$. Following state 0, $p_0'(\tau) = \lambda\bar{p}e^{-\lambda\tau} > 0$.

If $\bar{p} > 1/2$, $p_0(\tau)$ increases continuously from zero to $\bar{p}$ and crosses $1/2$ exactly once. Therefore, $v_0'(\tau) > 0$ before the crossing and $v_0'(\tau) < 0$ afterward. Solving

$$\bar{p}\left(1 - e^{-\lambda\tau_0^*}\right) = \frac{1}{2}$$

yields

$$\tau_0^* = \frac{1}{\lambda}\log\left(\frac{2\bar{p}}{2\bar{p} - 1}\right).$$

At this point, $v_0(\tau_0^*) = 1/4$. Following state 1, $p_1'(\tau) = -\lambda(1 - \bar{p})e^{-\lambda\tau} < 0$.

If $\bar{p} < 1/2$, $p_1(\tau)$ decreases continuously from one to $\bar{p}$ and crosses $1/2$ exactly once. Therefore, $v_1'(\tau) > 0$ before the crossing and $v_1'(\tau) < 0$ afterward. Solving

$$\bar{p} + (1 - \bar{p})e^{-\lambda\tau_1^*} = \frac{1}{2}$$

yields

$$\tau_1^* = \frac{1}{\lambda}\log\left(\frac{2(1 - \bar{p})}{1 - 2\bar{p}}\right).$$

Under stationarity, $\text{Var}(X_t) = \bar{p}(1 - \bar{p})$. Moreover, $\mathbb{E}[X_t \mid X_{t-\tau}] = \bar{p} + (X_{t-\tau} - \bar{p})e^{-\lambda\tau}$,

So $\text{Var}(\mathbb{E}[X_t \mid X_{t-\tau}]) = e^{-2\lambda\tau}\bar{p}(1 - \bar{p})$. The law of total variance then gives $\mathbb{E}[\text{Var}(X_t \mid X_{t-\tau})] = \bar{p}(1 - \bar{p})\left(1 - e^{-2\lambda\tau}\right)$, whose derivative is strictly positive. □

*A.4 Proof of Corollary 2.*

For the OU process, $V''(a) = -2\kappa\sigma^2 e^{-2\kappa a} < 0$, so $V$ is strictly concave. If $A_2$ is a mean-preserving spread of $A_1$, the definition of convex order implies $\mathbb{E}[V(A_2)] \leq \mathbb{E}[V(A_1)]$.

Taking $A_1 = \mathbb{E}[A_2]$ with probability one gives Jensen's inequality, $\mathbb{E}[V(A_2)] \leq V(\mathbb{E}[A_2])$. Strict concavity yields strict inequality for any nondegenerate spread. □

*A.5 Proof of Theorem 3*

Let $g(a)$ be a nonnegative measurable cost associated with age $a$. During a renewal cycle of length $T$, age increases from zero to $T$, so total age-dependent cost during the cycle is

$$G(T) = \int_0^T g\,(a)\,da.$$

The renewal-reward theorem gives long-run average cost

$$\overline{g} = \frac{\mathbb{E}\left[\int_0^T g\,(a)\,da\right]}{\mathbb{E}[T]}.$$

Setting $g(a) = a$ gives

$$\mathbb{E}[A] = \frac{\mathbb{E}[T^2]}{2\mathbb{E}[T]} = \frac{\mathbb{E}[T]}{2} + \frac{\mathrm{Var}(T)}{2\mathbb{E}[T]}.$$

For Brownian dynamics, set $g(a) = \sigma^2 a$. Then

$$\overline{V}_B = \frac{\sigma^2\mathbb{E}[T^2]}{2\mathbb{E}[T]} = \frac{\sigma^2}{2}\left[\mathbb{E}[T] + \frac{\mathrm{Var}(T)}{\mathbb{E}[T]}\right].$$

For the OU process, set $g(a) = \frac{\sigma^2}{2\kappa}(1 - e^{-2\kappa a})$. Integrating over a renewal cycle gives

$$\int_0^T g\,(a)\,da = \frac{\sigma^2}{2\kappa}\left[T - \frac{1 - e^{-2\kappa T}}{2\kappa}\right].$$

Dividing its expectation by $\mathbb{E}[T]$ yields the stated expression for $\overline{V}_{OU}$.

Finally, $G''(T) = g'(T) \geq 0$ whenever the age-cost function $g$ is increasing. Hence $G$ is convex. A mean-preserving spread in $T$ raises $\mathbb{E}[G(T)]$, while the denominator $\mathbb{E}[T]$ remains fixed. Both Brownian and OU conditional variance are strictly increasing in age, so a nondegenerate spread produces a strict increase. □

*A.6 Proof of Proposition 1.*

Under Assumption 5, linearity of $\pi$ in $X_t$ implies $\pi \mid \mathcal{F}_{t-\tau} \sim N(P + \beta\hat{X}_t, \beta^2 Var(X_t \mid \mathcal{F}_{t-\tau}))$.

*Certainty Equivalent.* Under conditional normality and CARA utility, the certainty equivalent of the payoff is

$$CE = P + \beta\hat{X}_t - \frac{\gamma\beta^2}{2} Var(X_t \mid \mathcal{F}_{t-\tau}).$$

*Participation Condition.* Suppose the decision maker requires a reservation certainty equivalent *C*. The participation condition $CE \geq C$ implies

$$P \geq C - \beta\hat{X}_t + \frac{\gamma\beta^2}{2} Var(X_t \mid \mathcal{F}_{t-\tau}).$$

The final term represents the additional compensation, $\Pi(\tau)$, required for bearing uncertainty created by delayed information. □

*A.7 Proof of Corollary 3.*

From Proposition 1, $\Pi(\tau) = \frac{\gamma\beta^2}{2} v(s, x; \tau)$. By Corollary 1,

$$\frac{\partial v(s, x; \tau)}{\partial \tau}\Big|_{\tau=0^+} = \sigma^2(x, s).$$

Therefore,

$$\frac{\partial \Pi(\tau)}{\partial \tau}\Big|_{\tau=0^+} = \frac{\gamma\beta^2}{2} \sigma^2(x, s) > 0.$$

Theorem 1 gives $\Pi(\tau) = \frac{\gamma\beta^2}{2} \sigma^2 \tau + O(\tau^2)$, so to first order,

$$\frac{\partial \Pi}{\partial \sigma^2} \approx \frac{\gamma\beta^2}{2} \tau > 0. \quad \square$$

*A.8 Proof of Corollary 4.*

*Because*

$$\frac{\partial CE(\tau)}{\partial \tau} = -\frac{\gamma\beta^2}{2}\frac{\partial}{\partial \tau}\mathrm{Var}(X_t \mid \mathcal{F}_{t-\tau}),$$

the result follows from Corollary 1 locally and from $V'(\tau) > 0$ globally in the OU benchmark. □

*A.9 Proof of Proposition 2.*

Conditional expected utility is $\mathbb{E}[u(\pi) \mid p] = -e^{-\gamma P}\left(1 - p + pe^{-\gamma\beta}\right)$. Equating this expression with $-e^{-\gamma CE(p)}$ gives $CE(p) = P - \frac{1}{\gamma}\log\left(1 - p + pe^{-\gamma\beta}\right)$. Because conditional expected payoff is $P + \beta p$, $\Pi(p) = \beta p + \frac{1}{\gamma}\log\left(1 - p + pe^{-\gamma\beta}\right)$. Differentiation gives

$$\Pi'(p) = \beta - \frac{1 - e^{-\gamma\beta}}{\gamma(1 - p + pe^{-\gamma\beta})}$$

and

$$\Pi''(p) = -\frac{\left(1 - e^{-\gamma\beta}\right)^2}{\gamma(1 - p + pe^{-\gamma\beta})^2} < 0.$$

Thus, $\Pi(p)$ is strictly concave. Solving $\Pi'(p) = 0$ gives

$$p^{\dagger} = \frac{1}{1 - e^{-\gamma\beta}} - \frac{1}{\gamma\beta}.$$

Since $\Pi(0) = \Pi(1) = 0$, this stationary point is the unique maximum. Expanding around $\gamma\beta = 0$

$$p^{\dagger} = \frac{1}{2} + \frac{\gamma\beta}{12} + O((\gamma\beta)^3).$$

The path $p_0(\tau)$ increases from zero to $\bar{p}$, so it crosses $p^{\dagger}$ if and only if $\bar{p} > p^{\dagger}$. The path $p_1(\tau)$ decreases from one to $\bar{p}$, so it crosses $p^{\dagger}$ if and only if $\bar{p} < p^{\dagger}$. The corresponding premiums are therefore hump-shaped under the stated conditions. □

*A.10 Proof of Corollary 5.*

Multiply the average-variance expressions in Corollary 2 and Theorem 3 by $\gamma\beta^2/2$. □

*A.11 Proof of Proposition 3.*

No intermediary with reservation price above $r_{(1)}$ can win, because $i^*$ can undercut its price. The winner cannot sustain a price above $r_{(2)}$, because the second-lowest-cost rival can profitably undercut it. A price below $r_{(2)}$ cannot be the limiting price, because $i^*$ can raise its price toward $r_{(2)}$ without inviting profitable undercutting. Hence the Bertrand limit is $r_{(2)}$. □

*A.12 Proof of Proposition 4.*

Differentiating the expected incremental cost gives $J'(\tau) = [1 - p(\tau)]c + p'(\tau)[L - c\tau]$, where $c = \gamma\beta^2\sigma^2/2$. Because

$$p'(\tau) = -\frac{\Delta}{2\sigma(T-\tau)^{3/2}}\phi\left(\frac{\Delta}{\sigma\sqrt{T-\tau}}\right) < 0,$$

the second term captures the protective effect of latency: a longer lag reduces the probability of reclassification. At $\tau = 0, J'(0) = (1 - p_0)c - |p'(0)|L$.

The stated condition therefore implies $J'(0) < 0$, so zero latency cannot minimize expected cost. As $\tau$ approaches $T$, both $p(\tau)$ and $p'(\tau)$ converge to zero, while $J'(\tau)$ converges to $c > 0$. Thus, $T$ cannot be optimal either. Since $J$ is continuous on $[0, T]$, it attains a minimum; because neither endpoint can minimize it, at least one minimizer lies in $(0, T)$. □